# Experimental realization of a high Curie temperature CoFeRuSn quaternary Heusler alloy for spintronic applications

Ravinder Kumar and Sachin Gupta*

*Department of Physics, Bennett University, Greater Noida 201310, India*

*Corresponding author email: sachin.gupta@bennett.edu.in

## Abstract

We synthesize CoFeRuSn equiatomic quaternary Heusler alloy using arc-melt technique and investigate its structural, magnetic and transport properties. The room temperature powder X-ray diffraction analysis reveals that CoFeRuSn crystallizes in cubic crystal structure with small amount of $DO_3$ - disorder. The field dependence of magnetization shows non-zero but small hysteresis and saturation behavior up to room temperature, indicating soft ferromagnetic nature of CoFeRuSn. The magnetic moment estimated from the magnetization data is found to be 4.15 $\mu_B$/*f.u.*, which is slightly less than the expected Slater-Pauling rule. The deviation in the value of experimentally observed moment from the theoretical value might be due to small disorder in the crystal. The low temperature fit to electrical resistivity data show absence of quadratic temperature dependence of resistivity, suggesting half-metallic behavior of CoFeRuSn. The high Curie temperature and possible half-metallic behavior of CoFeRuSn make it a highly promising candidate for room temperature spintronic applications.

**Keywords:** CoFeRuSn, Half metallic, Quaternary Heusler Alloy, Spintronics

---

# Introduction

Spintronics is a rapidly growing multidisciplinary field, which utilizes spin property of an electron in addition to its charge property, resulting in additional functionalities over conventional electronics.[1-2] To unlock the full potential of spintronics, theoretical prediction and experimental realization of new materials with multifunctional properties are very important. Half-metallic (HM) materials are among the best suited candidates for spintronic applications as these materials possess unique electronic structures. HM materials show metallic behavior for one spin sub-band (say spin up) while semiconducting/insulating behavior for other spin sub-band (spin down), resulting in ideally 100% spin polarization at the Fermi level.[3] The presence of gap at the Fermi level for one spin direction can have consequences on various physical properties. Theoretical predictions and experimental demonstrations revealed the Half-metallic property in materials from various classes.[4] Among these, Heusler alloys stand out for their ability to exhibit tunable electronic and magnetic properties, making them highly promising candidates for spintronic devices.[5] Recently, a new class of Heusler alloys, called quaternary Heusler alloys drawn significant attention due to their novel spintronic properties.[6-7-8] Equiatomic quaternary Heusler alloys (EQHAs) have general formula *XX'YZ*, where *X*, *X'* and *Y* are transition metal elements and *Z* is main group element.[9] Several quaternary Heusler alloys have been predicted theoretically and only some of them are experimentally realized. Most experimental reports focus on quaternary Heusler alloys based on 3*d* transition metals, with only a limited number involving 4*d* transition metals like Ru and Rh in EQHAs.[6] Moreover, it has been observed that Co-based EQHAs show comparatively high Curie temperature and large magnetic moment and therefore are preferred for various spintronic applications. From these observations, it becomes intriguing to investigate Co based quaternary Heusler alloys incorporating 4*d* transition metal elements. Recently, two independent research groups have made theoretical predictions regarding the half-metallic behavior of EQHA CoFeRuSn, however there is no experimental investigation conducted on this material yet.[10-11]

In this work, we synthesize CoFeRuSn and carry out X-ray diffraction (XRD), magnetic and magneto-transport measurements to study its structural, magnetic and transport properties. The structural analysis of XRD pattern suggests LiMgPdSn prototype structure. The magnetization data show ferromagnetic nature of CFRS at room temperature. Detailed transport measurements reveal half metallic behavior at low temperature, which disappears at high temperature. The field dependence of magnetoresistance shows asymmetric behavior at low temperature.

## Experimental details

Polycrystalline sample of CoFeRuSn (CFRS thereafter) was synthesized using an arc melt technique in presence of high purity argon atmosphere. The constituent elements; Co, Fe, Ru, and Sn (with minimum 99.99 % at. purity) with their stoichiometric proportion were melted in water cooled copper hearth. As cast sample was sealed in evacuated quartz tube and annealed for seven days at 850 °C followed by ice-water quenching to improve homogeneity. To check the crystal structure of CFRS, room temperature XRD pattern was recorded on Bruker D8 Advance diffractometer with Cu $K_\alpha$ radiation ($\lambda$ = 1.54 Å). FullProf Suite software was used to analyze the obtained XRD pattern. Magnetic measurements were performed using vibrating sample magnetometer (VSM) attached to physical property measurement system (PPMS) (Cryogenic Limited, UK) as a function of temperature and magnetic field. Electrical transport measurements were carried out on PPMS using four probe method with an applied current of 10 mA.

## Results and discussion

Electronic structures of quaternary Heusler alloys, *XX'YZ* are very sensitive to atomic disorder due to occupancy of random sites, which also have implication on its magnetic properties.[9] The XRD data can give an idea about the type and degree of atomic disorder in Heusler alloys.[12] For example, when all the atoms in a quaternary Heusler alloy occupy their specific Wyckoff positions, it results in fully ordered crystal structure and presence of (111) superlattice reflection in XRD pattern. Swapping of Wyckoff position between *Y* and *Z* atoms results in *B*2 disorder, which kills (111) superlattice reflection in XRD pattern. Mixing of all the atoms results in *A*2 disorder, killing both (111) and (200) reflections in XRD pattern. It is important to note that (400) and (220) reflections are always there in XRD pattern indicating crystallized Heusler alloy.[12] The Rietveld refinement of room temperature powder XRD pattern is shown in Fig. 1. It can be seen from Fig. 1 that all (200), (111), (220) and (400) reflections are present in XRD pattern, indicating LiMgPdSn prototype crystal structure with space group $F\bar{4}3m$ (number 216). However, it is worth noting that the intensity of (111) reflection in XRD pattern is higher than the (200) reflection, suggesting small amount of $DO_3$ type disorder in the crystal. In general, $BiF_3$ ($DO_3$) disorder is introduced in *XX'YZ* quaternary Heusler alloy via the exchange of *X* or *X'* with Y or X or *X'* with Z sites.[9-13-14] Similar results were also reported in other EQHA such as CoFeVSb attributed to the small *DO*$_3$ disorder in the alloy. The lattice parameter determined from the Rietveld refinement of the powder XRD data is found to be 6.11 Å.

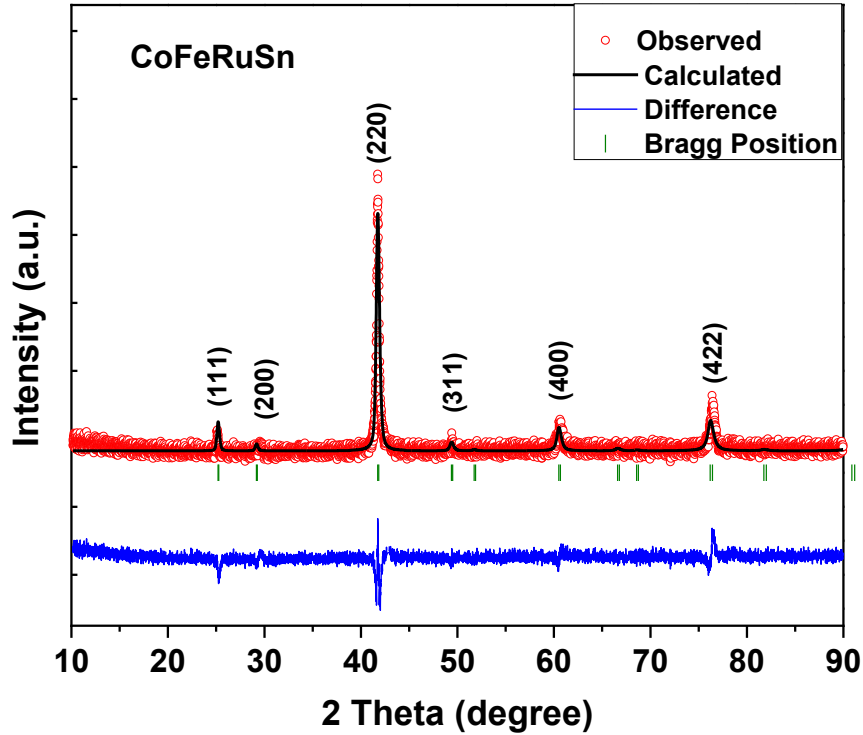

**FIG. 1.** The room temperature powder XRD pattern along with Rietveld refinement for CoFeRuSn. The bottom plot shows the difference between experimental and calculated patterns.

Fig. 2 shows the magnetic field, *H* dependence of magnetization, *M* as a function of temperature, *T*. The magnetic field was swept from + 90 kOe to – 90 kOe and the magnetization of sample was recorded at 4, 100 and 300 K. The inset shows field dependence of magnetization at smaller fields. It can be seen from the inset that CFRS shows small hysteresis at all the measured temperatures. The squarer shape of the *M-H* loop with small hysteresis and saturation behavior at 300 K indicates the soft ferromagnetic (FM) nature of CFRS. It is important to highlight the very weak field dependence of magnetization at higher fields, indicating a much higher Curie temperature of CFRS. It can be observed from Fig. 2 that both saturation magnetization and coercive field of CFRS decreases with increasing temperature. The value of saturation magnetization was found to be 4.15 $\mu_B/f.u.$ at 4 K. According to the Slater- Pauling (S-P) rule,[15-16] the total magnetic moment, $M$ of a quaternary Heusler alloy is directly related to the number of valence electrons, $N_v$ and can be estimated by the following equation;

$$M = (N_v-24)\ \mu_B/f.u. \qquad (1)$$

Since CFRS has 29 valence electrons, as per S-P rule, it is expected to have a magnetic moment of 5 $\mu_B/f.u.$ However, the experimental value of saturation moment determined from the magnetization data is 4.15 $\mu_B/f.u.$ at 4 K. The slight deviation in the magnetic moment from the theoretical expected value may be attributed to the anti-site ($DO_3$) disorder in the crystal as suggested from the XRD results. Similar results were also reported in many other QHAs such as CoFeVSb, CoRuVGa, NiFeMnGa, NiCoMnGa and CuCoMnGa.[14-17-18]

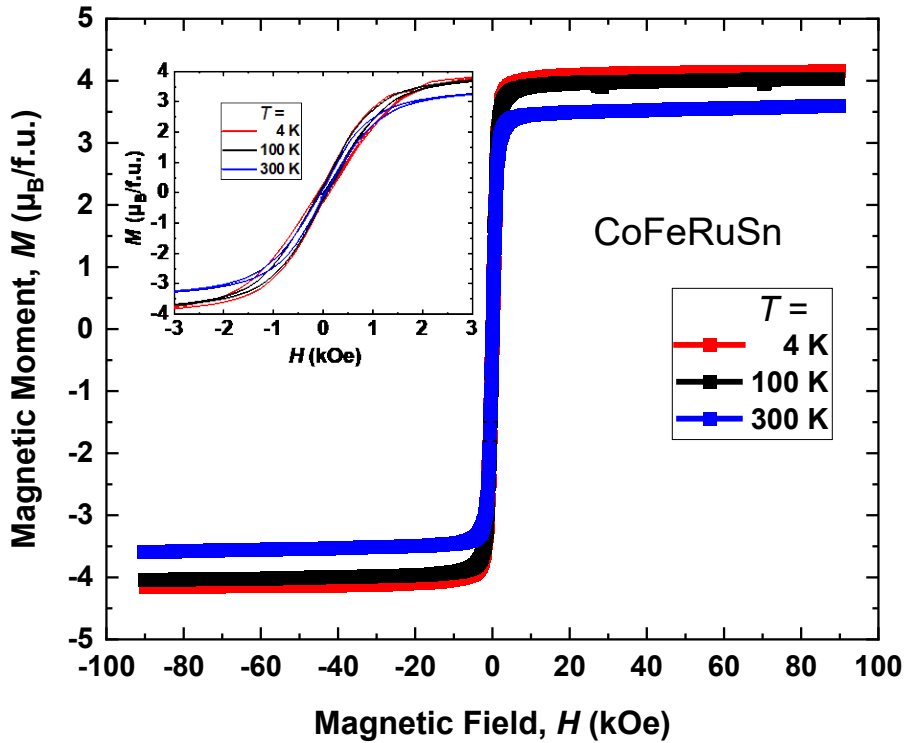

**FIG. 2.** The magnetic field, $H$ dependence of magnetization, $M$ as a function of temperature for CoFeRuSn. The inset shows magnetization at lower fields.

It has been observed that any changes in the electronic structure of a material are strongly reflected in the temperature dependence of electrical resistivity data. To investigate the electronic structure of CFRS, the zero-field electrical resistivity was measured in the temperature range of 2–300 K and is plotted in Fig. 3. The order of magnitude of the electrical resistivity is close to the values reported in other EQHAs.[17] The increase in the resistivity with temperature indicates the metallic behavior of CFRS. The comparatively large value of the residual resistivity ratio (RRR = $\rho_{300K}/\rho_{4K}$) of ~1.3 suggests least disorder and good quality of the sample.[9] In order to better understand the transport behavior of CFRS, the zero field resistivity data was fitted in two regimes: (i) low temperature (2-50 K) regime and (ii) high temperature (50-300 K) regime as shown in Fig. 3.

The temperature dependence of resistivity, $\rho(T)$ in low temperature regime (2-50 K) was fitted using the following equation.

$$\rho(T) = \rho_o + A\,T^n \tag{2}$$

Where $\rho_o$ is the residual resistivity and A is an arbitrary constant. The solid line in Fig. 3 shows fit to experimental data. Usually a quadratic temperature ($T^2$) dependence of resistivity has been observed in typical ferromagnets due to electron-magnon scattering.[19-22] $T^2$ dependence of resistivity is expected to be absent in the case of half-metals as the $T^2$ - term is related to single magnon scattering, which is forbidden due to existence of gap in minority spin channel.[23] The value of $n$ obtained from the fit of equation (2) is found to be 1.5, which indicates the suppression of electron magnon scattering due to absence of minority states at low temperature. Thus, non-quadratic dependence of low temperature resistivity data suggests possible half-metallic behavior in CFRS.

At high temperatures, the resistivity of a metal is usually dominated by the electron-phonon scattering, which results in almost linear behavior of resistivity at elevated temperature. In order to understand high temperature scenario, the temperature dependence of resistivity was fitted with the combined scattering contributions given as

$$\rho(T) = \rho_0 + \rho_{ph} + \rho_{mag} \tag{3}$$

Where $\rho_{ph}$, $\rho_{mag}$ are the electron-phonon scattering and magnonic scattering terms respectively. The temperature dependence of these scattering terms can be described as

$$\rho(T) = \rho_0 + BT + CT^2 \tag{4}$$

Where B, and C are arbitrary constants. The equation (4) fits well to experimental data as shown in Fig. 3 by solid blue line. The parameters estimated from the fit are given in Table 1.

Whether electron-phonon or electron-magnon scattering is significant, can be determined from the magnitude of the coefficient $C$. Electron- phonon scattering takes over when $C$ is small (order of $10^{-2}$ nΩcmK$^{-2}$). However, the electron-magnon scattering will have significant impact if $C$ is high (order of a few nΩcmK$^{-2}$).[24] It can be noted from the parameters given in Table 1 that high temperature resistivity is contributed by both electron-phonon scattering and one-magnon scattering terms. However, electron-phonon scattering is dominated over one-magnon scattering as magnitude of C is very small. Due to the existence of spin-flip scattering, half-metallic nature is lost at higher temperatures.

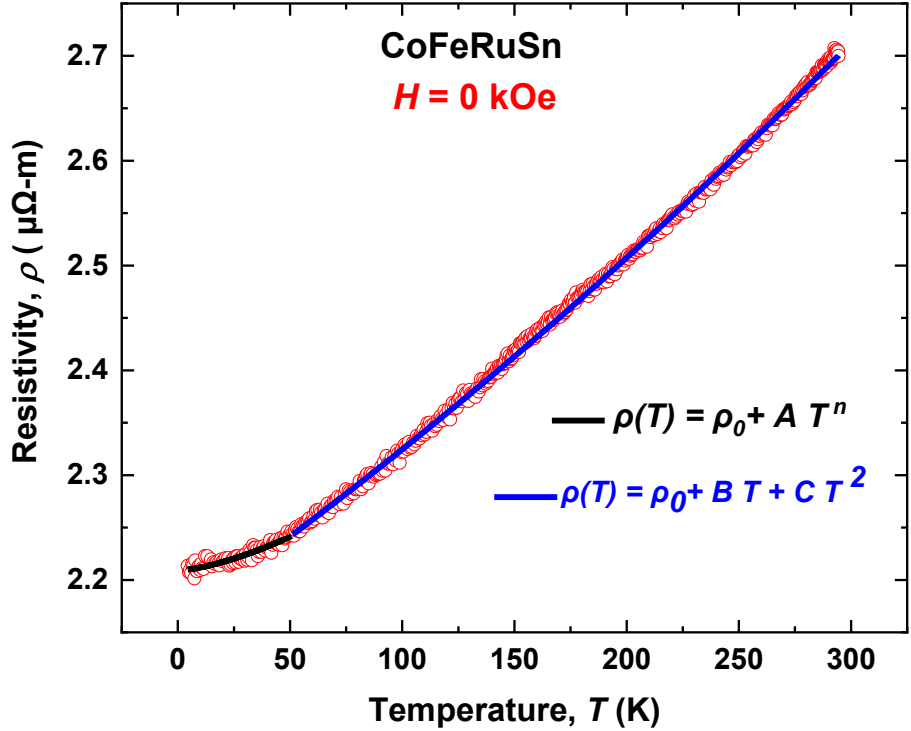

**FIG. 3.** The temperature dependence of electrical resistivity, $\rho$ for CoFeRuSn. The red circle represents the experimental data whereas black and blue solid lines represent low and high temperature fits to the experimental data.

**Table 1:** List of parameters obtained from low and high temperature electrical resistivity fit using equations (2) and (4) respectively.

| Material | Low temperature fitting parameters | | | High temperature fitting parameters | | |
|---|---|---|---|---|---|---|
| | A (µΩm K$^{-n}$) | n | Range (K) | B (µΩ m K$^{-1}$) | C (µΩ m K$^{-2}$) | Range (K) |
| CoFeRuSn | 9.0776×10$^{-5}$ | 1.5 | 2 - 50 | 0.0015 | 1.10268×10$^{-6}$ | 50 - 300 |

Fig. 4 shows the field dependence of magnetoresistance (MR) measured at 4 K in the the field range of ±50 kOe. The MR is defined as

$$MR = \frac{R_H - R_0}{R_0} \quad (5)$$

where $R_0$ and $R_H$ are the electrical resistances measured without and with field, respectively. It is clear from Fig. 4 that the magnitude of MR increases with increasing/decreasing field and does not show signature of saturation up to the field of 50 kOe. The magnitude of MR is small and shows asymmetric behavior with sweep up and down field. Similar behavior was also observed in CoFeVSb EQHA at low temperatures.[14]

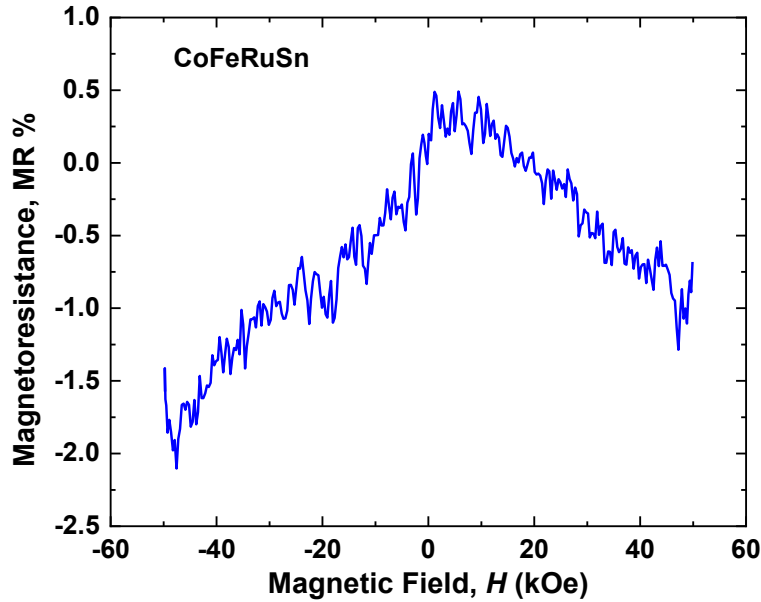

**FIG. 4.** The field dependence of magnetoresistance (MR) at 4 K for CoFeRuSn.

## Conclusions

We synthesize CoFeRuSn quaternary Heusler alloy using arc-melt technique. The XRD results show presence of superlattice reflections (111) and (200), indicating cubic structure with small amount of *DO$_3$* disorder. The magnetization measurements reveal soft ferromagnetic behavior at room temperature and a saturation moment of 4.15 $\mu_B / f.u$. The slight deviation of magnetic moment from the Slater-Pauling rule may be attributed to the *DO$_3$* disorder in the crystal. The electrical resistivity measurements suggest half metallic behavior at low temperature, which get disappeared at higher temperatures. The high Curie temperature and possible half metallic behavior of CoFeRuSn make this material very promising for room temperature spintronic applications.

## Acknowledgements

The authors thank Prof. K. G. Suresh, IIT Bombay, for providing access to the research facility for synthesizing the material and Mr. Barnabha for his helping hand. The authors also thank Prof. R. Chatterjee, IIT Delhi, for extending the research facility for preparing samples for transport measurements.